\documentclass[twocolumn,amsmath,amssymb,letterpaper]{revtex4}

\usepackage{graphicx}
\usepackage{dcolumn}
\usepackage{bm}

\begin{document}

 \title{Determining the Muon Mass in an Instructional Laboratory}
\author{Benjamin Brau}
 \email{bbrau@physics.umass.edu}
 \affiliation{Physics Department, University of Massachusetts, Amherst, MA 01003}
 \author{Christopher May, Robert Ormond, and John Essick}
 \email{jessick@reed.edu}
 \affiliation{Physics Department, Reed College, Portland, OR 97202}

\begin{abstract}
An instructional laboratory experiment to measure the muon mass is
described. Using coincidence-anticoincidence detection, the decay of a
cosmic-ray muon into an electron (or positron) is observed in a multiplate
spark chamber, and recorded with a triggered CCD detector. The energy of the
charged decay-product particle is then quantified by counting the number of
spark gaps $n_S$ it traverses before being stopped by the chamber's aluminum
plates. By running this apparatus under computer-control for several hours,
the number of product-particles $N(n_S)$ with various $n_s$-values is
obtained. The muon mass is obtained by a least-squares fit, which compares
the experimentally observed $N(n_S)$ with simulation values predicted for
this distribution by the Fermi description of muon decay via the weak
interaction. We present the results for the muon mass we have obtained and
discuss the simulation we developed to account for the observed skewing of
$N(n_S)$ due to the various directions the spark-producing product particles
move as well as the escape of some of the higher-energy particles from the
chamber.
\end{abstract}

\maketitle

\section{Introduction}

Earth is continually bombarded by cosmic rays, high-speed subatomic
particles produced by astrophysical processes such as supernovae. Many of
the first experiments in elementary particle physics used cosmic rays as
their particle source, and even today, they continue to be studied in
experiments that probe the very highest particle energies. While this source
of high-speed particles is omnipresent and available at no cost, the flux of
cosmic rays is relatively low. Thus experiments done with cosmic rays can
best probe processes with high intrinsic rates.

For over forty years, the presence of high-speed muons at Earth's surface
has allowed educators to design instructional laboratory experiments that
illustrate the concerns and experimental methods of elementary particle
physics. These muons are the byproduct of collisions between cosmic rays and
nuclei near the top of Earth's atmosphere. By far, the most widely adopted
such experiment measures the muon's lifetime
\cite{Melissinos,Hall,Owens,Lewis,Ward,Coan}, but others include observation
of the time dilation effect \cite{Coan,Easwar} and determination of the
muon's magnetic moment. \cite{Amsler} In 1964, researchers at the University
of Michigan suggested that the mass of the muon could be measured through an
experiment that analyzed tracks produced by muon decay within a spark
chamber. \cite{UMich} These researchers built a prototype system and
demonstrated the feasibility of this approach. However, at that time, the
necessary setup required too much specialized technical expertise for most
educators to copy and the experiment was too cumbersome to perform because
track images were captured on photographic film. Thus, this experiment has
never been embraced by instructional lab developers. In this paper, we
describe the muon mass experiment and how it can be implemented today with
relative ease through the use of modern instrumentation, including CCD-based
image acquisition. We go on to show how an accurate value for the muon mass
can be determined from the data gleaned from this experiment using an easy
to replicate software simulation program.
\section{Theory}

A muon is a negatively charged elementary particle from the lepton family
with a mass that is about 200 times that of an electron. With effectively
100\% probability, a free muon decays into an electron, an electron
antineutrino, and a muon neutrino with a mean lifetime of 2.2 $\mu$s. That
is, the muon's principal decay mode is $\mu^- \rightarrow e^- + \overline
\nu_e + \nu_\mu$. Similarly, the principal decay mode of the muon's
antiparticle is $\mu^+ \rightarrow e^+ + \nu_e + \overline \nu_\mu$.  As a
result of the interaction of cosmic rays with the nuclei of air molecules at
the top of Earth's atmosphere, Earth's surface is bombarded by a stream of
high-speed charged secondary cosmic ray particles, 75\% of which are a
roughly equal mixture of muons and antimuons. In the vertical direction, the
total flux per unit solid angle of these secondary particles is about ${\rm
0.66/cm^2 \cdot sr \cdot min}$. \cite{Melissinos} Using this value, along
with the empirically determined cosine-squared variation of particle flux
with angle from the vertical, it can be shown that the flux of charged
secondary cosmic-ray particles impinging on a horizontally aligned detector
is on the order of ${\rm 1/cm^2 \cdot min}$.

The goal of our experiment is to stop a significant number of secondary
cosmic-ray muons within a spark chamber and observe each decay into an
electron (to save words, we will simply describe the decay of a muon into an
electron, but our discussion likewise applies to the decay of an antimuon
decay into a positron). The spark chamber is a charged-particle detector
that makes visible the muon's path prior to being stopped as well as the
path of the product electron after the decay. Two product neutrinos are also
produced by the muon's decay, but because they are neutral their paths are
not recorded by the spark chamber. Our chamber consists of a vertical stack
of aluminum plates. Most cosmic-ray muons that enter the chamber have
energies greater than 1 GeV and so, although slowed by the plates, will pass
completely through the chamber. However, a small population of low-energy
muons will be effectively ``stopped" (i.e., slowed to kinetic energies on
the order of the aluminum's ionization energy) as they pass through the
sequence of plates, suppressing the time-dilation effect that allowed these
short-lived particles to traverse the height of the atmosphere. Roughly half
of these ``stopped" muons then decay freely via $\mu^- \rightarrow e^- +
\overline \nu_e + \nu_\mu$, while the other half are captured by atoms and
undergo the process of inverse beta decay $\mu^- + \rm Al \rightarrow e^- +
\nu_\mu + \rm Mg^*$ (which is undetected by the spark chamber). On the other
hand, all of the ÒstoppedÓ antimuons decay freely via $\mu^+ \rightarrow e^+
+ \nu_e + \overline \nu_\mu$.

Consider the decay of a  stationary muon of mass $m_\mu c^2$ (consistent with the shorthand used in the field of particle physics, we will call this quantity the ``muon mass," although it is actually the ``muon rest energy") into an electron, an electron antineutrino, and a muon neutrino. Applying conservation of energy to this process, we find $m_\mu c^2=E_e+ E_{\overline \nu_{e}} + E_{\nu_\mu}$, where the energies $E_e$, $E_{\overline\nu_e}$, and $E_{\nu_\mu}$ of the three product particles can be taken (to a good approximation) to be entirely kinetic energy because each of these particles has a rest mass much less than $m_\mu c^2$. Additionally, applying conservation of momentum, we get $0=\vec p_e+\vec p_{\overline \nu_e}+\vec p_{\nu_{\mu}}$, where $\vec p_e$, $\vec p_{\overline \nu_e}$, and $\vec p_{\nu_\mu}$ are the relativistic momenta of the product particles. With three product particles, the kinetic energy of the electron is not uniquely determined by the conservation laws, but instead a range of  $E_e$-values is possible. The minimum value of this range is $E_e=0$ (neglecting the electron's mass) in which case the electron is at rest and the two neutrinos move oppositely directed with equal momentum magnitude. At the other extreme, the maximum electron energy occurs when the two neutrinos both move in the opposite direction to the electron. Then, assuming the product particles are all massless, it is easy to show from the conservation laws that $E_e=\frac{1}{2} m_\mu c^2$. Thus, the allowed range of electron energies is $0\leq E_e \leq \frac{1}{2} m_\mu c^2$. The Fermi description of muon decay via the weak interaction, \cite{Griffiths} in which the decay (in lowest order) is mediated by a $W$ boson of mass $M_w$, yields the following relation for the probability per unit time $d\Gamma$ that a product electron will be produced with energy in the interval $E_e$ to  $E_e+dE_e$
\begin{multline}\label{1}
\frac{d\Gamma}{dE_e}=\bigg{(}\frac{g_w}{M_w c^2}\bigg{)}^4 \frac{1}{384\pi^3\hslash } \bigg{(}{m_\mu c^2 E_e}\bigg{)}^2 \bigg{(}3-4 \frac {E_e}{m_\mu c^2}\bigg{)} \\ 0 \leq E_e \leq \frac{1}{2} m_\mu c^2 
\end{multline}
where $g_w$ is the weak coupling constant. Higher order electromagnetic
effects modify this expression slightly (by approximately 1\%). Noting that
$d\Gamma/dE_e$ is independent of angle, Eq. ({\ref 1}) predicts that, within
a given time interval and solid angle, the probability $P(E_e)$ of product
electrons being produced with energy $E_e$ is
\begin{equation}\label{2}
P(E_e)=C(m_\mu c^2 E_e)^2 (3-4 E_e/m_\mu c^2)
\end{equation}
where $C$ is a constant. Hence, by fitting an experimentally determined plot of $P(E_e)$ vs. $E_e$ to this predicted functional form, a value for $m_\mu c^2$ can be determined.

\section{System Overview}

\begin{figure}
\includegraphics[width=0.5\textwidth]{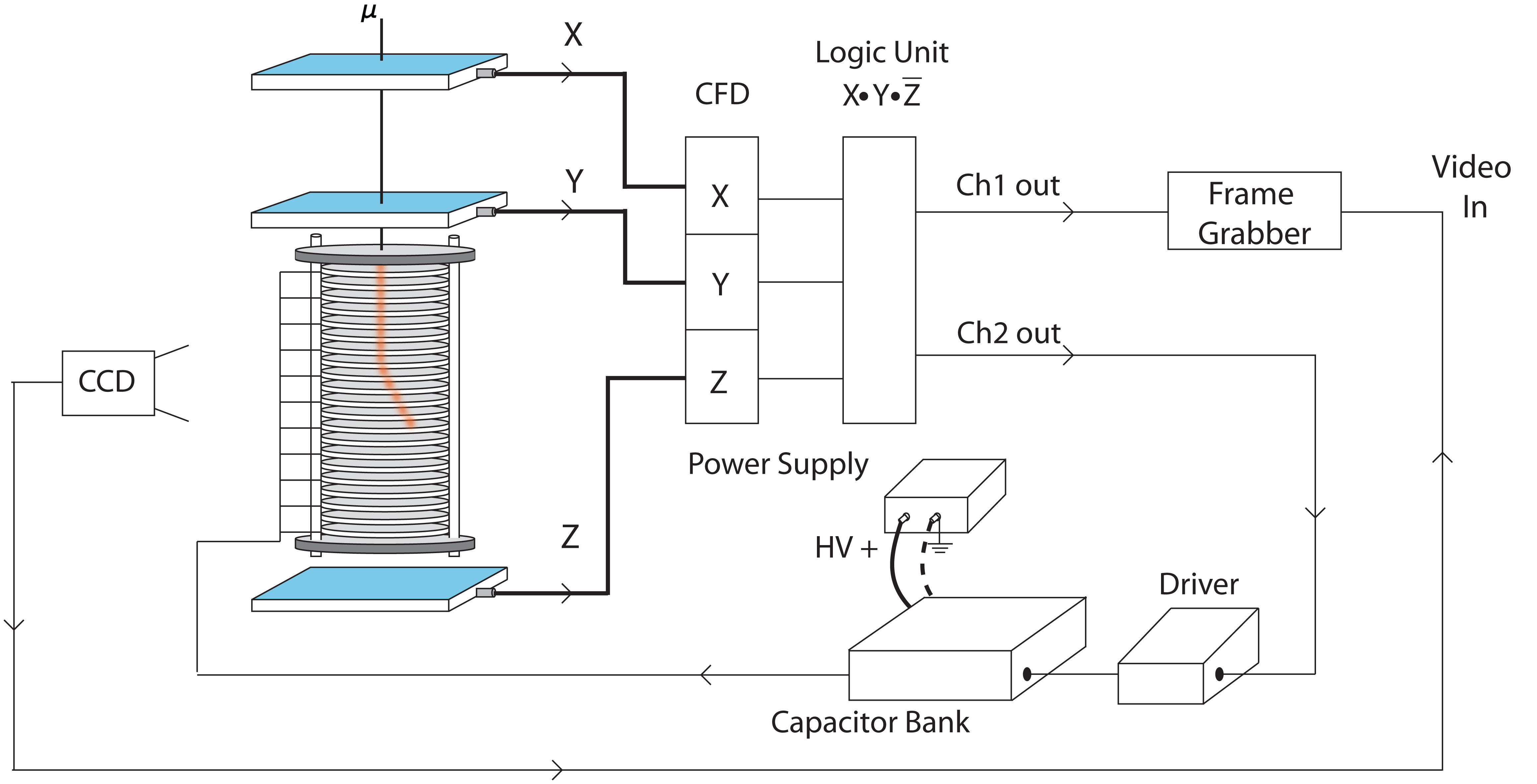}
\caption{Experimental set-up, which includes multiplate spark chamber, coincidence/anticoincidence electronics to detect muon decay within chamber, fast high-voltage pulser to create spark trail, and triggered CCD-based image acquisition to store data.}
\label{setup}
\end{figure}

Our system consists of four main components: spark chamber, particle-decay detection electronics, high-voltage pulsing electronics, and CCD-based image acquisition (see Fig. \ref{setup}). A short description of each of these components follows; more detailed descriptions are given in the Appendix.

The spark chamber consists of a stack of 21 disk-shaped aluminum plates. Each plate has thickness $t_p=0.9525 \ {\rm cm} \ (=3/8 \ {\rm inch})$ and radius $R=7.62 \ {\rm cm} \ (=3 \ {\rm inch})$ and is separated from its neighboring plates by ``spark gaps" of thickness $t_g=0.635 \ {\rm cm} \ (=1/4 \ {\rm inch})$ filled with a noble gas (in our case, neon). The odd-numbered plates are electrically grounded, while the even-numbered plates are connected to a fast high-voltage, high-current pulser. When a charged particle moves within the chamber, it ionizes the noble gas along its path, leaving behind a trail of ionized atoms and liberated electrons. If the pulser can be triggered to apply sufficiently high voltage (typically 6000 V) to the even-numbered plates more quickly than the mean ion-electron recombination time (on order of $5 \  \mu \mathrm{s}$), sparks will form in the gaps between the plates along the low-resistance ionized track, making visible the trajectory of the moving charged particle.

The decay of a cosmic-ray muon within the spark chamber is detected through the use of three scintillation detectors stacked vertically about the chamber. Each detector consists of a horizontally aligned light-tight 20 cm $\times$ 20 cm $\times$ 2 cm plastic scintillator coupled to a blue-light sensitive photomultiplier tube (PMT). When a high-speed charged muon passes through such a detector, it causes the scintillator material to fluoresce \cite{pdg} and the resulting photons are detected and converted to a short ($\approx$ 5 ns), negative-going electrical signal by the attached PMT. This signal is amplified and then passed through a discriminator which converts the PMT signal into a short digital pulse. Two of these detectors (henceforth called $X$ and $Y$) are placed above the chamber, separated by a vertical distance of approximately 1 meter, while the third detector ($Z$) is placed slightly below the chamber. When a muon with (close to) normal incidence enters the chamber from above, and then subsequently decays within the chamber, digital pulses will be produced by $X$ and $Y$, while a digital pulse will not be produced by $Z$, assuming the decay-product electron is stopped within the chamber. Thus, if the digital outputs of the three discriminators are input to a logic unit that performs the coincidence/anticoincidence operation $X \cdot Y \cdot \overline{Z}$ ($X$ AND $Y$ AND not$Z$), the output of this logic unit signals the muon's decay within the spark chamber.

The $X \cdot Y \cdot \overline{Z}$ output of the logic unit is connected to the input of a high-speed pulser (``driver"). This driver unit produces the proper voltage pulse to trigger a high-voltage, high-current switch (``thyratron"), which then causes a bank of capacitors to discharge, applying a (negative) pulse of several thousand Volts to the even-numbered plates of the spark chamber.

Finally, a video camera focused on the spark chamber is configured to send its analog video signal to the video input of a frame grabber board within the expansion slot of a PC, while the $X \cdot Y \cdot \overline{Z}$ output of the logic unit is connected to the board's trigger input. The frame grabber is controlled via a LabVIEW program, which acquires and stores a single image of the chamber whenever the logic unit detects a muon decay event.

\section{Experimental Procedure}

With high voltage applied to all three scintillation detector PMTs, each PMT output (after passing through a preamplifier) is input to a separate channel of a quad constant fraction discriminator (CFD). Using an oscilloscope, the width of NIM digital pulses output by each CFD channel (in response to PMT input signal pulses) is adjusted to be 300 ns. In our setup, each detector's cross-sectional area $A=20 \ \mathrm{ cm} \times 20 \ \mathrm{ cm}=400 \ \mathrm{\ cm}^2$, so the rate of digital pulses produced by secondary cosmic ray particles on each channel is about $(1/\mathrm{ cm}^2 \cdot \mathrm{min})(400 \ \mathrm{ cm}^2)(\mathrm{1 min/60 \ s}) \approx \mathrm{7/s}$. 

The digital-pulse outputs of the CFD's three channels $X$, $Y$, and $Z$ are input into the logic unit. The TTL output of this logic unit is connected to the high-speed driver, which triggers the spark chamber operation. Proper functioning of the setup can be verified by programming the logic unit to perform the logic operation $X \cdot Y$, so that the spark chamber is triggered for any charged particle (i.e., not just those that decay) that passes into the chamber through the top two detectors. The expected rate of such $X \cdot Y$ coincidence is found as follows: $X$ and $Y$ each have  area $A=400 \mathrm{\ cm}^2$, and are separated by distance $d=100 \mathrm{\ cm}$. Thus, at each point on $Y$, the solid angle subtended by $X$ is $\Delta \Omega \approx \frac{400 \ \mathrm{ cm}^2}{(100 \  \mathrm{ cm})^2}=0.04 \ \mathrm{sr}$, and so the expected rate at which vertically downward secondary cosmic ray particles (75 \% of which are muons and antimuons) will pass through both detectors producing coincidence signals is $(0.66\mathrm{/cm}^2 \cdot \mathrm{sr} \cdot \mathrm{min})(400 \ \mathrm{cm}^2)(0.04 \  \mathrm{sr})=10\mathrm{/min}$.

To execute a data run, the logic unit is configured to perform the logic operation $X \cdot Y \cdot \overline{Z}$, the video camera is focused on the spark chamber along with its reflection in a nearby mirror angled at $45^\circ$ (to obtain an orthogonal viewpoint of the spark tracks), and a LabVIEW-based image acquisition program is started. Each time the logic unit triggers the spark chamber, it also sends a TTL trigger pulse that triggers the frame grabber board to acquire the video camera's current frame and save it as a jpeg file on the computer's hard drive with a unique file name. The experiment is run under computer control in a darkened room overnight, typically resulting in about 100 saved images per hour. 

\begin{figure}
\includegraphics[width=0.4\textwidth]{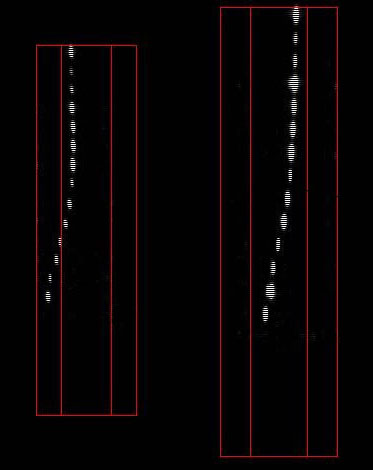}
\caption{Typical spark image of a muon decay with the direct and orthogonal (created by 45$^\circ$-angled mirror) view of the chamber on right and left, respectively. Here, the muon enters from the top and, after it decays, the product electron moves leftward and toward the back of the chamber while producing 6 sparks. Superimposed on each view is a software-generated outer and inner rectangle defining the boundaries of the entire chamber and the half-radius volume, respectively. The upper portion of the half-radius volume is the fiducial volume within which a valid decay must initiate.}
\label{sparks}
\end{figure}

After a data run, all the images are inspected. Fig. \ref{sparks} shows a
sample image of a muon decay event, which consists of a direct and an
orthogonal view of the chamber. The orthogonal view is the image produced by
a $45^\circ$-angled mirror located at the left side of the chamber; this
image is smaller than the direct view because it is farther from the camera
than the chamber. In Fig. \ref{sparks}, the muon enters from the top of the
chamber and, after it decays, the product-electron moves downward to the
left and toward the back of the chamber. To analyze this decay image, we
construct two lines, one delineating the path of the incoming muon, and the
other marking the path of the product electron. The kink at the intersection
of these lines defines the location at which the muon decayed. Once the
point of decay is established, one then counts the number of sparks $n_S$
created by the electron as it is brought to rest (or else escapes from the
chamber) during its traverse through the aluminum plates. In
Fig. \ref{sparks}, $n_S=6$.

In contrast to the image in Fig. \ref{sparks}, many of our acquired images do not clearly show a muon decay event for the following reasons. First, although from Eq. (\ref{1}), we expect an isotropic distribution of product-electron tracks, i.e., as many images with an electron going up as down from the decay site, we never get a clear decay image showing an upward moving electron. This observation is simply an artifact of our spark-chamber detection method; to record a decay event with an upward moving electron would require each spark gap above the decay site to produce two sparks, one for the downward incident muon and another for the upward directed electron. Unfortunately, each spark gap can only produce a single spark. Second, for a downward directed electron, we find that a spark is created only if the electron's path deviates up to roughly $30^ \circ$ from the chamber's axis. This observation results from the fact that the electric field produced by the plates is along the chamber's axis (i.e., normal to the plates) and if the ionized trail is inclined too steeply from this direction, no spark forms. Thus, in our inspection process, we select out the images which clearly display muon decay events. Because of the details of spark formation described above, these images always have downward directed product-electrons, moving at some angle less than about  $30^ \circ$ to the chamber's axis.

\begin{figure}
\includegraphics[width=0.45\textwidth]{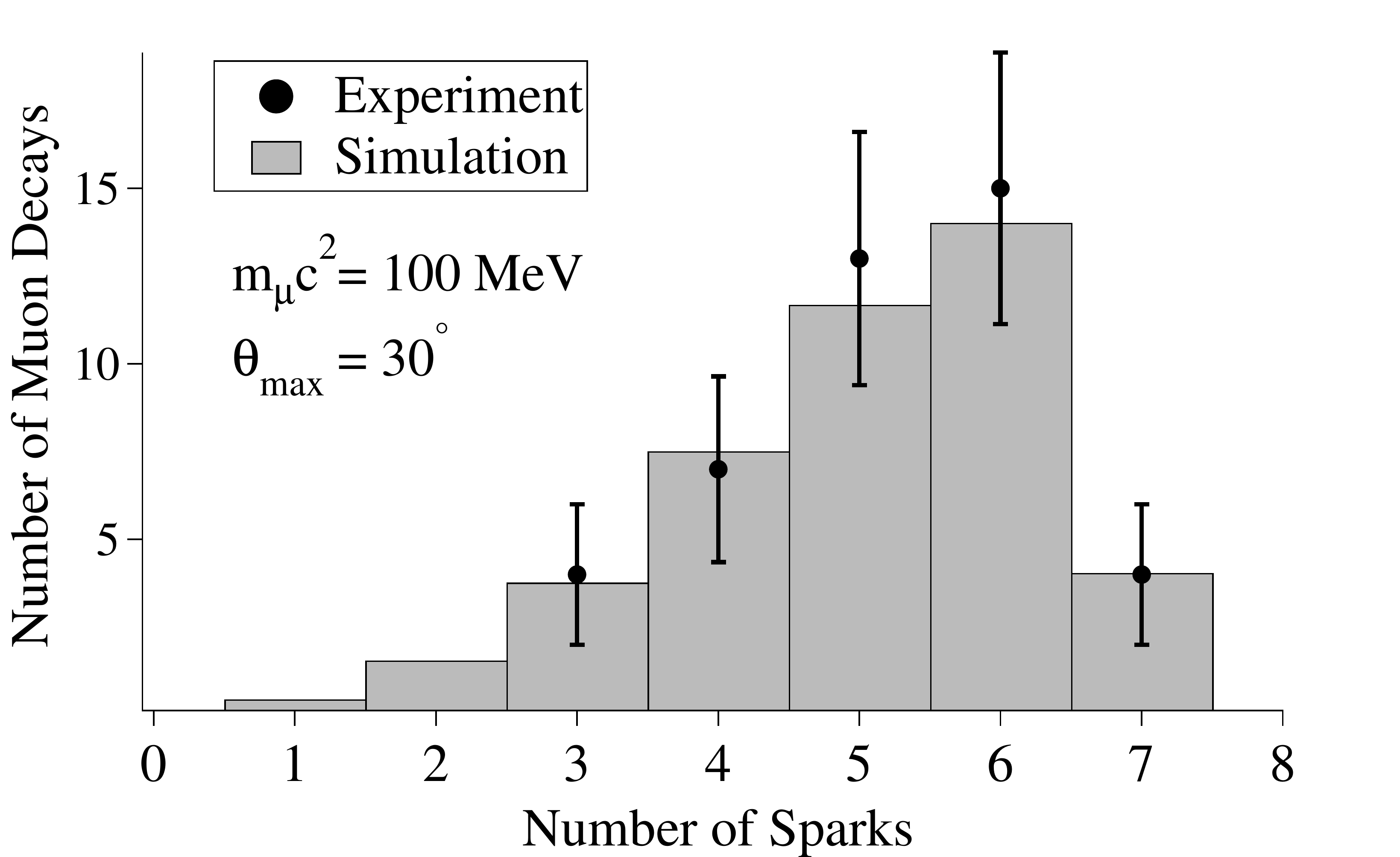}
\caption{Comparison of experimental data ($N_{exp}^{tot}=43$) with the best-fit result of the simulation, assuming  $m_\mu c^2 = 100$ MeV and $\theta_{max}=30^\circ$. Error bars on the data are computed as the square root of the associated count number.}
\label{data_1}
\end{figure}

For reasons delineated in the next section, we analyze only the subset of those images for which the muon decay occurs within the chamber's ``fiducial volume," which we define to be a cylinder in the top half of the chamber with a radius $R_{fid}=R/2$ measured from the chamber's central axis. After this selection process, one is left with a total of $N_{exp}^{tot}$ images that record muon decays (typically $N_{exp}^{tot} \approx 50$). Using these data, we then determine the number of muon decays $N_{exp}(n_S)$ in which product electrons produce $n_S$ sparks, and then plot $N_{exp}(n_S)$ vs. $n_S$ as in Fig. \ref{data_1}.

\section{Analysis of Data}

The product electron's initial kinetic energy $E_e$ (i.e., its energy immediately after the muon decay) can be determined by analyzing its path through the chamber's sequence of aluminum plates and spark gaps. Assuming a ballpark value of $m_\mu c^2 \approx 100$ MeV,  Eq. (\ref 1) indicates that $E_e$ will be in the range from 0 to 50 MeV. As electrons with initial energies in this range are slowed during their travel through aluminum, the two most significant energy-loss processes are ionization and bremsstrahlung,\cite{Leo} whose stopping powers $S=-dE/dl$, i.e., energy loss per unit length, are approximately a constant $S_\circ$ and proportional to instantaneous energy $E$, respectively. In particular, the total stopping power for an electron in aluminum \cite{Katz,Alum} is given by 
\begin{equation} \label{3}
-\frac{dE}{dl}=S_{ioniz}+S_{brem}= S_\circ+ \frac{E}{X_\circ}
\end{equation}
where $S_\circ = \rm {5.09\ MeV/cm}$ and $X_\circ =8.9 \ {\rm cm}$ is the ``radiation length" for aluminum. Note that $S_{ioniz}=S_{brem}$ at the ``critical energy" $E_{crit}=S_\circ X_\circ= 45 \rm\ {MeV}$, and that for electron energies much less (much greater) than this critical energy, ionization (bremsstrahlung) is the dominant energy-loss process. Since the electron energies in our experiment do not deviate greatly from $E_{crit}$, we must account for both energy-loss mechanisms; thankfully, it is easy to show from Eq. (\ref 3) that the path length $l$ in aluminum required to bring an electron with initial energy $E_e$ to rest is 

\begin{equation} \label{4}
l=X_\circ \ln\bigg{[}1+\frac{E_e}{E_{crit}} \bigg{]}
\end{equation}

The ideal experimental procedure would be as follows: For each observed muon decay, measure the length $l$ of the product electron's path through the chamber's aluminum plates, and then use Eq. (\ref {4}) to calculate the electron's initial energy $E_e$. Repeating this process for all the observed decays, determine the probability $P(E_e)$ of a product particle being produced with energy $E_e$, then fit the resulting plot of $P(E_e)$ vs. $E_e$ to Eq. (\ref 2), and extract a value for the muon mass from the fit. Because we obtain two perpendicular views of the electron's path, in principle, we could carry out this ideal procedure. However, in practice, it proves very difficult to determine the length of the electron's path in three dimensions from our data, and so we take another tack.

As mentioned in the previous section, we simply characterize the path of a product electron by counting the number of sparks $n_S$ it creates while moving in the chamber. Clearly, there is not a one-to-one correspondence between $n_S$ and $E_e$, e.g., with the same initial energy $E_e$, an electron passing through the aluminum plates at normal incidence will produce more sparks than one passing through the plates at a large incident angle (because of the increased path length per plate).  However, by writing a straightforward Monte Carlo simulation, we can account for all of the possible $n_S$ an electron with $E_e$ can produce and so predict the distribution of $n_S$, which can then be compared with the experimentally determined distribution.

One issue that we must accurately address in our simulation is the sub-population of product electrons that escape the chamber before they are completely stopped (e.g., electrons created near the chamber's edge and moving at large angles relative to the chamber's axis). Because the number of sparks created by an escaping electron is quite sensitive to its direction of travel, to model this type of event accurately requires our simulation to explore a large parameter space. To minimize the influence of escaping electrons on the outcome of our simulation, we restrict our data set in the following way so that this sub-population of electrons is as small as possible: First, we note from Eq. (\ref 4) that the most energetic ($\approx$ 50 MeV) product electron will require about 7 cm of aluminum to be stopped. Since each plate is approximately 1 cm thick, about 7 plates will be required to stop such an electron that is directed straight downward. Also, we find empirically that a spark is created only if the electron's path is inclined up to roughly $30^ \circ$ relative to the chamber's axis. Thus, by considering muon decays that occur only with the cylindrical ``fiducial volume"  defined to be in upper half of chamber with radius $R_{fid}$ equal to half the plate radius $R$, most of the spark-producing electrons will interact with enough aluminum to be stopped before escaping the chamber.

\begin{figure}
\includegraphics[width=0.45\textwidth]{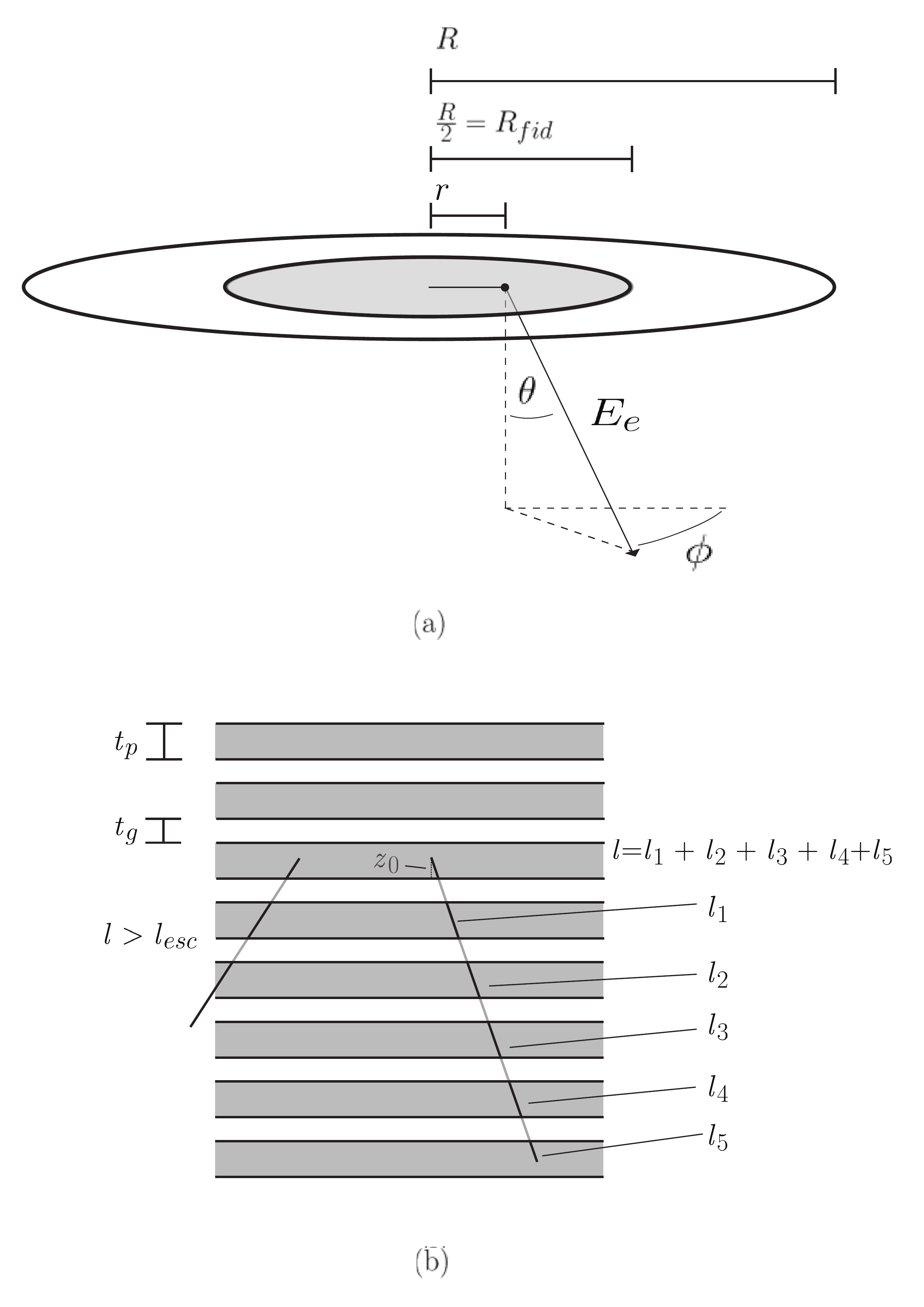}
\caption{Geometry of simulation. (a) Initial position $r$ and angles $\theta$ and $\phi$ for product electron; and (b) length $l$ traversed within aluminum for stopped and escaping electrons.}
\label{geometry}
\end{figure}

For our simulation,\cite {Sim} because charged particles lose negligible energy when passing through the chamber's spark gaps, we assume each muon and resulting product electron are stopped within the interior of a plate. At each muon-decay site, the product electron is assumed to emanate downward, located a radial distance $r$ ($<R_{fid}$) from the chamber's axis and directed at polar angle $\theta$ relative to the chamber's (downward) axis and azimuthal angle $\phi$ as in Fig. \ref{geometry}. If the decay occurs a (perpendicular) distance $z_\circ$ above the plate's bottom surface and the electron's path length for being stopped in aluminum is $l$, then the number of sparks produced is given by

\begin{equation} \label{5}
n_S=1+{\rm floor} \bigg{[} \frac{l \cos \theta -z_\circ}{t_p} \bigg{]}
\end{equation}
 where the floor function return the highest integer less than or equal to its argument. Also, if $l>l_{esc}$, the electron will escape the chamber before being stopped, where

\begin{equation} \label{6}
l_{esc} \approx  \frac{-r \cos \phi + \sqrt {r^2 \cos^2 \phi+(R^2-r^2)}}{\sin \theta} \cdot  \frac{t_p}{t_p + t_g}
\end{equation}
 
Hence, the algorithm for our simulation is as follows: Assume a value for $m_\mu c^2$. For the product electron emanating from a muon-decay site, randomly assign  $z_\circ$, $r$, $\theta$ and $\phi$ to be within the range 0---$t_p$, 0---$R/2$, 0 --- $\theta_{max}$ and 0 --- $2\pi$, respectively, and also randomly assign the electron to have an initial kinetic energy $E_e$ within the range of 0 --- $m_\mu c^2/2$. Then, using Eq. (\ref 4) and (\ref 6), we calculate $l$ and $l_{esc}$. If $l>l_{esc}$, the electron's path length in the chamber is set to $l=l_{esc}$. Finally, the number of sparks produced by this electron is found from Eq. ({\ref 5}) and, based on Eq. (\ref 2), the number of times this event is expected to occur is assigned the value

\begin{equation}\label{7}
N_{sim}(n_S)=Dr(m_\mu c^2 E_e)^2 (3-4 E_e/m_\mu c^2)
\end{equation}
where $D$ is a normalization constant. The factor of $r$ is included because the differential area at radius $r$ is $dA=2\pi r dr$ and so the number of muon decays at differing radii is expected to be proportional to a geometric scaling factor of $r$. After iterating this procedure $N_{sim}^{tot}$ times (typically $N_{sim}^{tot}=10000$), the accumulated $N(n_S)$ values are normalized so that $\sum_{n_S}N_{sim}(n_S)=N_{exp}^{tot}$, where $N_{exp}^{tot}$ is the total number of muon decays observed experimentally. Then, the predicted distribution $N_{sim}(n_S)$ is compared with the experimentally determined $N_{exp}(n_S)$  as shown in Fig. \ref{data_1}.

\begin{figure}
\includegraphics[width=0.45\textwidth]{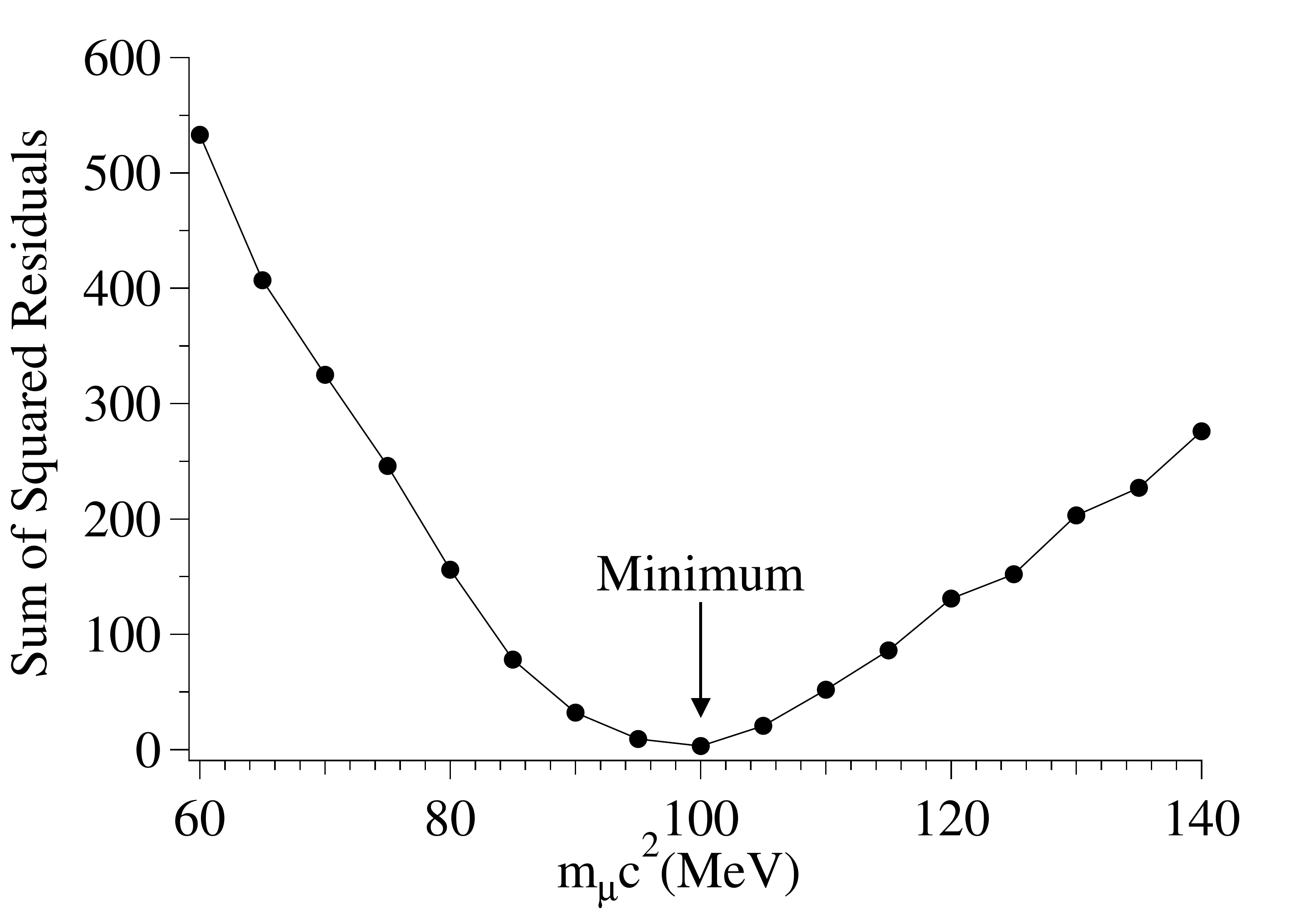}
\caption{Sum $S$ of squared residuals vs. muon mass $m_\mu c^2$, assuming $\theta_{max}=30^\circ$. Minimum occurs are 100 MeV.}
\label{squares}
\end{figure}

Finally, we iterate our simulation several times, where we assume a different value for $m_\mu c^2$ each iteration. At the conclusion of each iteration, we first compare the experimentally determined $N_{exp}(n_S)$  with the predicted $N_{sim}(n_S)$ by calculating the residuals $r(n_S)=N_{exp}(n_S)-N_{sim}(n_S)$, and then determine the sum $S$ of the squared residuals 
\begin{equation}
S=\sum_{n_S}\bigg{[}N_{exp}(n_S)-N_{sim}(n_S)\bigg{]}^2
\end{equation}
After concluding all of the iterations, we plot $S$ vs. $m_\mu c^2$, and then obtain our ``least-squares best-fit"  value for the muon mass from the minimum of this plot. As shown in Fig. \ref{squares}, the best-fit value derived from the data given in Fig. \ref{data_1} is $m_\mu c^2=100 \ \rm{MeV}$. A weighted least-squares calculation yields the same result.

\begin{figure}
\includegraphics[width=0.45\textwidth]{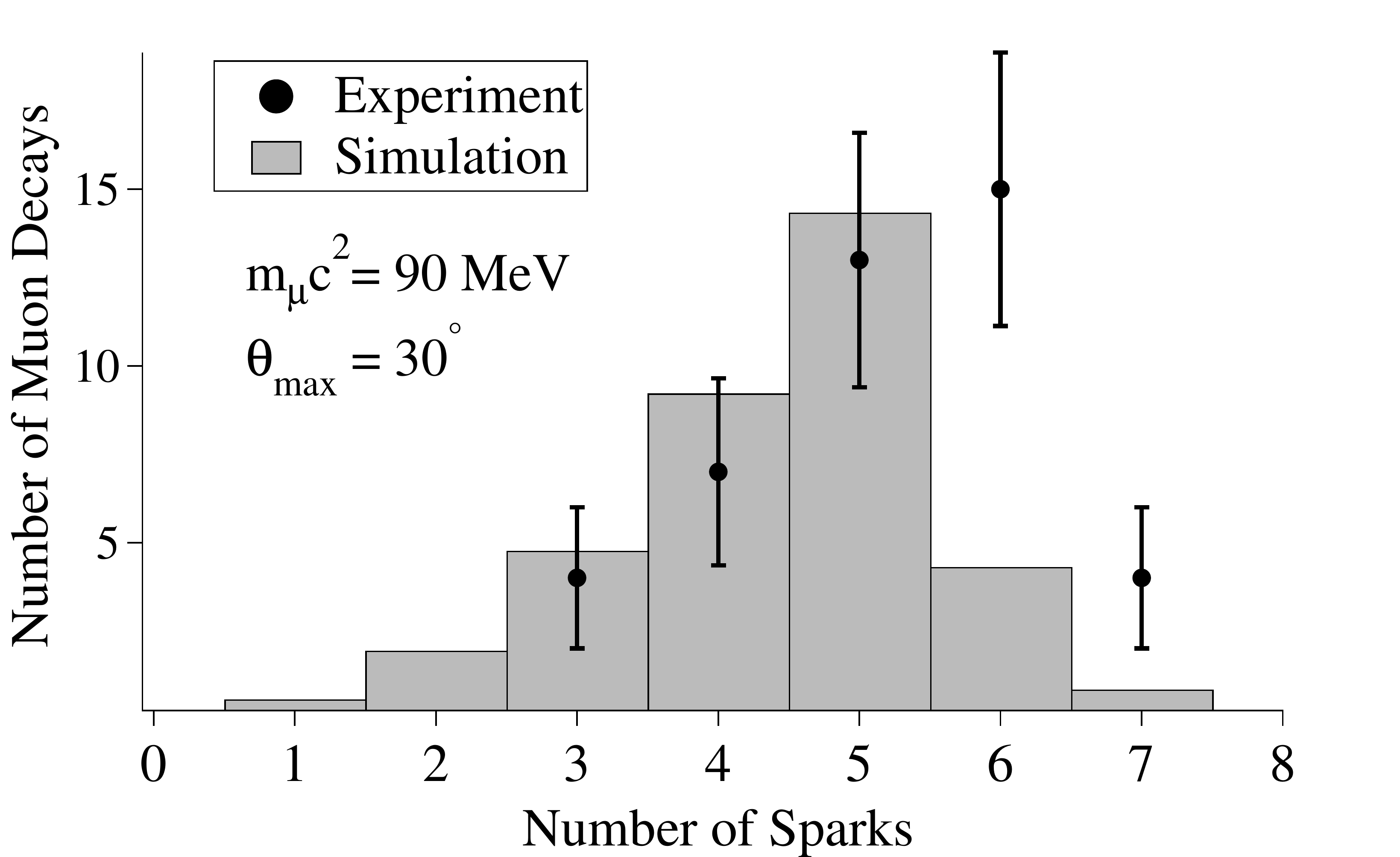}
\caption{Comparison of experimental data ($N_{exp}^{tot}=43$) with result of simulation, assuming  $m_\mu c^2 = 90$ MeV and $\theta_{max}=30^\circ$.}
\label{data_2}
\end{figure}

\section{Conclusion}
As shown in Fig. \ref{data_1}, when our simulation is run under the assumption that $m_\mu c^2 =100$ MeV and $\theta_{max}=30^\circ$, $N_{exp}(n_S)$ and $N_{sim}(n_S)$ are in very good agreement. To demonstrate the sensitivity of our approach to the assumed value of the muon mass, Fig. \ref{data_2} displays the result of simply changing the value of $m_\mu c^2$ used in the simulation to 90 MeV. We see that $N_{exp}(n_S)$ and $N_{sim}(n_S)$ now significantly disagree. Alternatively, if the plot of $S$ vs. $m_\mu c^2$ is generated under the assumption of  $\theta_{max}=40^\circ$, the plot's minimum occurs at $m_\mu c^2=$105 MeV. Through such investigations as well as from noting that the minimum region in Fig. \ref{squares} is relatively flat over the region from 95 to 105 MeV, we estimate our experimental value for $m_\mu c^2$ to be (100$\pm$5) MeV, a range consistent with the accepted muon mass value of 106 MeV.

\section{Appendix: Construction Details}

Spark Chamber: The design of the spark chamber is an adaptation of a spark chamber constructed by A.M. Sachs in the mid-1960s.\cite{Sachs} The chamber is constructed  of 3/8$''$ thick, round aluminum plates. The chamber consists of 20 gaps and 21 plates. The 19 inner plates are 6$''$ in diameter; the top and bottom plates are 7$''$ in diameter. The conductive aluminum plates of the chamber are separated by 6$''$ outer diameter, 5 1/4$''$ inner diameter plexiglass spacer rings, 1/4$''$  in thickness. 
It is essential that the chamber be completely leak tight, and O-rings provide a simple cost effective means of sealing the chamber. Therefore, the gas seals between the aluminum plates and plexiglass spacers are made with O-rings. The plates of the chamber have grooves to accommodate 1/8$''$  thick, 5 1/2$''$ inner diameter O-rings. The 19 inner plates have 1/4$''$ holes to allow gas to flow freely in the chamber. The chamber is held together by four threaded rods; these rods are electrically shielded from the pulsed plates with 1/2$''$ O.D., 1/4$''$ I.D. plexiglass sleeves.
 
The chamber is filled with pure neon at slightly over atmospheric pressure (approximately 1.1 atm) to ensure that any leakage in the chamber will be outward, not inward. This prevents any atmospheric gas impurities from leaking into the chamber. Electronegative gas impurities such as oxygen have detrimental effects on the chamber's operation.

Scintillation Detectors: Each scintillation detector  consists of a light-tight 20 cm $\times$ 20 cm $\times$ 2 cm plastic scintillator (Saint-Gobain BC-408) coupled to a blue-light sensitive photomultiplier tube (Hamamatsu R7400). The detectors were purchased as assembled units (Saint-Gobain 8X8.8BC408/.5L-X). The PMTs of all three detectors are connected in parallel to a high-voltage supply (Stanford Research Systems PS310) and operated at 1000 V. The signal output of each PMT is passed through a fast (1 ns rise time) noninverting preamplifier (Ortec VT120C) with fixed 20$\times$ gain. 

Coincidence/Anti-Coincidence Electronics: Each of the amplified PMT signals from the $X$, $Y$, and $Z$ scintillation detectors is sent to an input of a multi-channel constant fraction discriminator (CFD) with shaping delay set to 4 ns. To avoid incorrect transient logic states, cable delays from detector to CFD are arranged to ensure that $Z$ arrives to the CFD before $X$ and $Y$. We use the quad 200 MHz Ortec 935, a NIM-bin module whose negative-NIM logic output pulse widths are adjustable (typical value in our experiment is 300 ns). The $X$, $Y$, and $Z$ NIM logic output signals from the CFD are input to each of two separate channels of a quad 4-Input Logic Unit (Ortec CO4020), where each channel is configured to perform the logical operation $X \cdot Y \cdot \overline{Z}$. When this condition is met, each channel outputs a TTL pulse with width of 800 ns, one of which triggers the high-speed driver, while the other triggers the frame grabber board.

\begin{figure}
\includegraphics[width=0.5\textwidth]{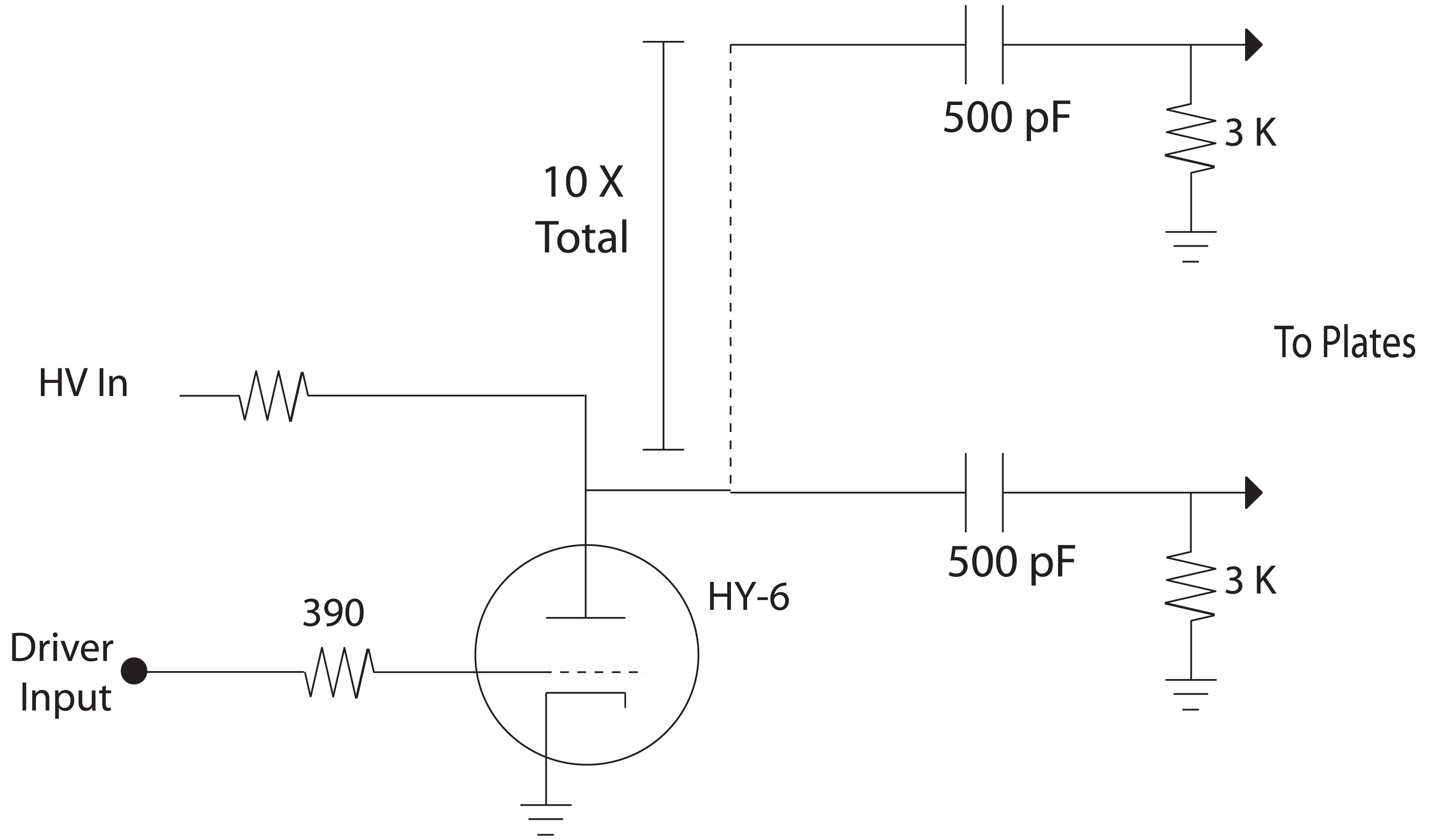}
\caption{Circuit diagram for capacitor bank.}
\label{capacitorbank}
\end{figure}

High-Voltage Pulsing Electronics: Triggerable high-voltage pulses are applied to the spark chamber plates using the circuit shown in Fig. \ref{capacitorbank}. Inside a homemade ``capacitor bank," several thousand (typically 6000) Volts from a 10 kV high-voltage supply (Bertan 230-10R) are applied to the anode of a hydrogen thyratron (Perkin Elmer HY-6) as well as the ``high-voltage" side of each $C=500$ pF capacitor (Sprague 20 DK-T5) in a parallel array of capacitors. The ``low-voltage" side of each capacitor is connected to a unique plate of the spark chamber as well as a current limiting $R=3 \ \mathrm{M} \Omega$ resistor, whose opposite end is grounded.  This circuit is triggered by a commercially available Thyratron Driver (Perkin Elmer TM-27, modified to trigger on 1.5 V).\cite{Star} When the Driver receives a TTL pulse from the Logic Unit, it outputs a 2-$\mu$s wide pulse of amplitude 800 V with a rise time of less than 150 ns. This trigger pulse is passed into the homemade pulser box, where it is applied to the grid of the hydrogen thyratron, driving the thyratron into conduction for 2 $\mu$s and causing the ``low-voltage" sides of the capacitors to drop to large negative voltage for a time period of approximately $\tau = RC = (3 \ \mathrm{M} \Omega )(500 \mathrm{\ pF}) = 1.5 \mathrm{ms}$.

Image Acquisition System: A monochrome progressive scan CCD camera head (Sony XC-55) with zoom lens (Navitar Zoom 7010) or else a camcorder (Canon ZR 40) with its autofocus option turned off is focused on the spark chamber directly as well as its perpendicular reflection in a mirror angled at $45^{\circ}$. With the camera's shutter speed at 1/30 s, the analog video signal is sent to a frame grabber board (National Instruments PCI-1405) plugged into a PCI expansion slot of a PC. The TTL output from the Logic Unit is connected the frame grabberÕs TRIGGER input. Under the control of a LabVIEW based program, when each TRIGGER pulse is received, the current frame acquired by the camera is stored under a unique filename as a jpeg file on the PC's hard drive.

\begin{acknowledgments}
We would like to thank Greg Eibel for invaluable assistance in machining our chamber and to Jim Brau and David Griffiths for helpful discussions. 
\end{acknowledgments}


\begin{thebibliography}{}
\bibitem[Sim(2008)]{Sim} Here we present a simplified simulation algorithm, which can be duplicated by anyone with programming skill in some computer language. This algorithm is based on a ``research-grade" simulation one of us [BB] wrote in Root, the standard analysis framework used in contemporary high-energy physics. We have found that the simplified simulation produces effectively the same results as our Root-based simulation.
\end{thebibliography}
\end{document}